# HAlign-II: efficient ultra-large multiple sequence alignment and phylogenetic tree reconstruction with distributed and parallel computing

Shixiang Wan[1]*, Quan Zou[1]*

1. School of Computer Science and Technology, Tianjin University, Tianjin, China

* Email: shixiangwan@gmail.com, zouquan@tju.edu.cn

## Abstract

Multiple sequence alignment (MSA) plays a key role in biological sequence analyses, especially in phylogenetic tree construction. Extreme increase in next-generation sequencing results in shortage of efficient ultra-large biological sequence alignment approaches for coping with different sequence types. Distributed and parallel computing represents a crucial technique for accelerating ultra-large sequence analyses. Based on HAlign and Spark distributed computing system, we implement a highly cost-efficient and time-efficient HAlign-II tool to address ultra-large multiple biological sequence alignment and phylogenetic tree construction. After comparing with most available state-of-the-art methods, our experimental results indicate the following: 1) HAlign-II can efficiently carry out MSA and construct phylogenetic trees with ultra-large biological sequences; 2) HAlign-II shows extremely high memory efficiency and scales well with increases in computing resource; 3) HAlign-II provides a user-friendly web server based on our distributed computing infrastructure. HAlign-II with open-source codes and datasets was established at http://lab.malab.cn/soft/halign.

**Introduction**

Multiple sequence alignment (MSA) is a necessary step for analyzing biological sequence structures and functions, phylogenetic inferences, and other basic fields in bioinformatics [1]. Given the rapid increment of biological sequences in next-generation sequencing [2], difficulty arises from insufficiency of available state-of-the-art methods for addressing ultra-large sources.

Increasingly more different parallelization strategies are implemented for reducing time and space complexity of MSA. These strategies can be mainly categorized into three levels: multiple threads based on central processing unit (CPU) on a single machine, multiple threads based on graphics processing unit (GPU) on a single machine, and multiple threads based on CPUs or GPUs on cluster machines. CPU-based multiple threads, which are common and effortless, suit small-scale sequence alignment. With emergence of bottlenecks in increasing clock frequency of multi-core CPUs, Moore's law became meaningless [3]. Based on NVIDIA GPU, compute unified device architecture (CUDA) technique was designed for efficient parallelism [4]. GPU functions in real-time rendering of screens, because hundreds of cores in GPUs can efficiently calculate pixels or coordinates in parallel. However, under limited video memory size and bandwidth, alignment of ultra-large sequences becomes difficult or even impossible [5]. With high computational cost, most naive algorithms attempted to reduce time and space complexity to cope with ultra-large analysis tasks.

Recently, large-scale distributed computing was applied extensively to various biological analyses, such as ClustalW-MPI [6], Hadoop-BAM [7], HAlign [8], and HPTree [9]. For next-generation sequencing, CloudDOE [10], BioPig [11], and SeqPig [12] were implemented; these software benefited from using open-source distributed frameworks. Different from traditional single machine systems, distributed computing systems perform load-balancing for fault-tolerant parallelized tasks and can be easily extended to cheaper devices for improvement of computing power. Additionally, distributed computing systems based on MapReduce framework present more abstract interfaces and more elastic computing resources than those based on Message Passing Interface (MPI) [13]. Ultra-large biological sequence analysis can be efficiently addressed by assembling distributed and parallel computing systems with numerous cheap devices [14-16].

Although HAlign software, which is based on Hadoop framework [17], exhibits better computing power and expansibility than other strategies running on a single machine. Apache Spark framework works up to 100 times faster than Hadoop, especially in iterative operators. Apache Spark can also accelerate real-world data analytics approximately 40 times faster than Hadoop and can even be employed to scan one TB data in five- to seven-second latency [18]. Based on Spark framework [19], Marek et al. developed SparkSeq [20], which can be used to analyze nucleotide sequence with considerable scalability. Zhao et al. developed SparkSW [3], which can carry out Smith-Waterman algorithm [21] in load-balancing way on a distributed system to cope with increasing sizes of biological sequence databases. However,

SparkSeq can only work with nucleotide sequences but not with protein sequences; thus, Smith-Waterman algorithm in SparkSW cannot achieve peer performance on nucleotide sequences, and it is not user-friendly. Both SparkSeq and SparkSW are fairly suitable for developers, they do not support generation of phylogenetic trees.

We implement HAlign-II based on HAlign work, HPTree work, and Apache Spark framework to address ultra-large multiple biological sequence alignment and to construct phylogenetic trees with rapid growth of biological sequence database. HAlign-II shows extremely high memory efficiency, which is efficient for MSA and phylogenetic trees construction, scales extremely well with increases in computing resources, and provides a user-friendly web server that is deployed on our infrastructure.

The rest of this paper is organized as follows. In the following section, we first introduce the Apache Spark framework. Based on Spark framework, we describe in detail Smith-Waterman algorithm for protein sequence alignment, trie trees algorithm for similar nucleotide sequence alignment, and neighbor-joining (NJ) method [22] for phylogenetic trees construction. Thereafter, we present datasets and comparative experiments with state-of-the-art tools and evaluate memory efficiency and scalability of our method. Last, preceding experimental results are discussed, and conclusion of the study is provided.

## Methods

### Overview of Apache Spark

Apache Hadoop and Apache Spark are famous open-source frameworks in the field of distributed computing. Hadoop mainly contains Hadoop Distributed File System (HDFS) [17] for distributed storage and MapReduce programming model for big datasets [23]. HDFS stores data on inexpensive machines, providing dependable fault-tolerant mechanism and high-aggregate bandwidth across clusters. Spark aims to blueprint a programming model that extends applications of MapReduce model and achieves high computational efficiency-based memory cache.

(Figure 1)

Spark designs an abstract data structure named resilient distributed datasets (RDDs) [18] to support efficient computing and to ensure distribution of datasets on cluster machines. RDDs support extensive variety of iterative algorithms, a highly efficient SQL engine Shark, and a large-scale graph computing engine GraphX. RDDs staying in memory cache will visibly reduce load time when requiring replication, especially in iterative operations. From Figure 1, to further reduce time and cost, two types of operations in RDDs are designed: transforms and actions [18]. Transforms only deliver computing graphs, which only describe how to compute and not how to carry out computing operations, such as map and filter operation. Actions carry out computing, such as reduce and collect operations, results of which are stored as new RDDs. Based on these operations, RDDs are efficiently executed in parallel. To ensure dependable fault tolerance, RDDs will be recomputed after data loss, for example, because of halting of individual machines. Based on RDDs, Spark can implement up to 100 times theoretical speed than Hadoop in real-world datasets [18].

**Smith-Waterman algorithm for protein sequences with Spark**

With its high sensitivity, Smith-Waterman algorithm [22] can locally align object and subject sequences to obtain similarity segments based on dynamic programming; however, global alignment results cannot be obtained. In the past decades, this algorithm was cited over 8,000 times in the biological field.

Smith-Waterman algorithm can search the best alignment location through given scoring methods, such as substitution matrix and gap-scoring scheme. Negative scoring matrix cells of this algorithm are set to zero, which is necessary for achieving alignment location. Traceback procedure of alignment starts from highest scoring matrix cell and proceeds until a cell with score of zero is encountered, thereby yielding the highest local alignment scoring. Suppose that $n$ and $m$ correspond to respective lengths of $A$ and $B$ sequences, then substitution matrix and gap-scoring scheme are respectively represented by $s(a,b)$ and $W_k$. Then, Smith–Waterman algorithm creates scoring matrix H and initials the first row and column; the process can be formulated as follows:

$$H_{k0} = H_{0l} = 0, (0 \leq k \leq n, 0 \leq l \leq m). \tag{1}$$

Then, the rest of matrix H should be filled with similarity scores, which are formulated as follows:

$$H_{ij} = max \begin{cases} H_{i-1,j-1} + s(a_i, b_j), \\ max_{k \geq 1}\{H_{i-k,j} - W_k\}, \\ max_{l \geq 1}\{H_{i,j-l} - W_l\}, \\ 0 \end{cases} \quad (1 \leq i \leq n, 1 \leq j \leq m). \tag{2}$$

where $H_{i-1,j-1} + s(a_i, b_j)$ represents similarity scores between $a_i$ and $b_j$, $H_{i-k,j} - W_k$ corresponds to matched scores when $a_i$ points to the end of a $k$ length gap, $H_{i,j-l} - W_l$ is the matched scores when $b_j$ points to the end of a $l$ length gap, and 0 indicates absence of similarity.

Figure 2 shows gradual traceback from the highest-score matrix cell to lowest-score matrix cell, looping to dynamic programming based on zero-score matrix cell. The algorithm obtains inserted space positions and generates pairwise alignment results.

(Figure 2)

As high time and space complexity of Smith-Waterman algorithm poses challenges concerning ultra-large datasets, this paper implements this algorithm on distributed computing system based on Spark framework.

As shown in Figure 3, the entire processing procedure is partitioned into two MapReduce steps. In the first step, the extracted center star sequence based on Smith–Waterman algorithm becomes a broadcast variable to align other sequences for filling inserted space matrix cells; this sequence records positions and numbers of inserted space. Then, first reduction generates the last and longest center star sequence for further calculations. Score matrix and center star sequence are cached in memory, spreading the center star sequence to each data node. Next, final pairwise alignment is initiated by inserted space matrix and each individual sequence. Finally, HDFS stores MSA results.

(Figure 3)

**Trie trees method for similar nucleotide sequences with Spark**

Smith-Waterman algorithm is accurate and mature and thus is suitable for protein sequence alignment of more complex structures and elements (for example, the 20 kinds of amino acids in humans). However, to obtain high similarity of most nucleotide sequences during alignment, time complexity of Smith-Waterman algorithm extremely increases, especially with ultra-large nucleotide sequences. Hence, this work considers tree-based data structures to address the problem in ultra-large nucleotide sequence alignment. Based on tree data structures, a series of multiple sequence alignment methods are availably applied; such methods include BLAT [24] and Hobbes [25]. According to HAlign [8], trie tree serves as an efficient data structure for storing multiple sequences; this structure quickly indexes common sub-strings from long strings and accelerates MSA search. A trie tree only features one root node and $n$ leafs for $n$ nucleotide sequences [26]. Additionally, trie tree can speed up search in linear running time by failure links.

Two primary steps can be used to realize MSA using trie tree: select a center star sequence for pairwise alignment and to integrate inserted spaces. Center star sequence contains the most segments among all sequences, thereby implying that it is the most similar to other sequences. As large-scale nucleotide sequences are similar, the first sequence represents the center sequence. Thereafter, other sequences are aligned to center sequence based on unmatched segments from the trie tree. In HAlign-II, this step is designed as numerous highly parallel operations across data construction of RDDs and is partitioned into memory on multiple workers. Pairwise alignment costs linear running time instead of exponential running time. Suppose that $n$ similar nucleotide sequences with average length of $m$ exists. Then, time complexity of trie tree algorithm is $O(n^2 m)$; trie tree algorithm requires less running time than the original center star method (time complexity is $O(n^2m^2)$). For $n-1$ times pairwise sequence alignment, time complexity is $O(nm^2)$. However, practical time consumed is far less than theoretical value because matched segments are skipped in high sequences. If $n \ll m$, then practical time consumed can be regarded as linear. In the last step, multiple alignment results are partitioned into new RDDs and delivered to multiple distributed workers for calculation. Center star sequence and its alignment results spread to entire Spark cluster as shared similar constants, as presented in Figure 3, to further reduce running time.

**NJ method for constructing phylogenetic trees with Spark**

Phylogenetic trees can be built using distance-based, maximum parsimony, and maximum likelihood approaches [9]. NJ approach [22] represents one of the distance-based approaches, and according to HPTree work, it is time-efficient and suitable for ultra-large sequences data.

As shown in Figure 4, based on parallel computing, we first cluster all MSA results into several clusters. Then, we calculate individual phylogenetic tree based on individual clusters. Last, all phylogenetic trees are merged on clusters into the final evolution tree. The approach comprises two key steps: initial clustering and MSA. MSA methods are determined by Smith-Waterman algorithm for protein sequences and trie trees algorithm for similar nucleotide sequences. Then, we highlight the initial clustering procedure. Approximately 10% of all sequences are selected by random sampling for initial clustering. Then, functional distance of each pairwise sequence is calculated, clustered, and labeled until all sequences are identified. When few clusters whose number of elements is over 10%, then they are merged into other clusters; otherwise, they are divided into more balanced clusters until balanced construction. The entire procedure is designed for Spark parallel model.

(Figure 4)

## Results and Experiments

### Datasets and experimental environment

For protein sequence alignment, BAliBASE [27] is regarded as golden benchmark, with BAliBASE 4 as the newest version. We employ the newest and largest R10 data sets $\Phi_{Protein}$ as our protein sequence datasets. Based on protein alignment database, a series of state-of-the-art methods can be used as compared objects. Additionally, we use human mitochondrial genomes $\Phi_{DNA}$ and 16s rRNA $\Phi_{RNA}$ as nucleotide sequences datasets [28], which are utilized for comparison with previous HAlign work [8]. After MSA, phylogenetic trees are generated by MSA results on Spark. Table 1 shows more detailed information regarding biological datasets.

(Table 1)

The main research object of HAlign-II includes ultra-large biological sequences. We use average sum-of-pairs (avg SP) score, which was proposed by Zou et al., as performance metric to evaluate results of MSA [8]. SP represents the sum of pairwise alignment scores. However, evaluation of large-scale data cases requires more intuitive average SP score. In pairwise alignment, one score is added when two nucleotides differ, and two scores are allotted when a space is inserted; otherwise, no score is added. Then, we consider maximum likelihood value as performance metric to evaluate results of ultra-large phylogenetic tree [9].

As HAlign-II contains three types of biological sequence alignment and phylogenetic tree construction based on Spark distributed system, our experimental environment consists of a cluster comprising 12 workstations. Each workstation features 384 GB physical memory with Intel Xeon E5-2620 processors, and each processor contains eight processing cores. Based on Ubuntu 16.04 operating system and Spark 2.0.2, a series of experiments are presented in succeeding sections.

**Comparison with state-of-the-art tools**

We select a series of state-of-the-art tools to compare with HAlign-II and evaluate its performance on addressing ultra-large datasets. Our comparison eliminates KAlign method [29], which is incompletely suitable for large-scale datasets. Similarly, phangorn [30], RAxML [31], and STELLS [32] are eliminated because of their nearly intolerable time consumption. As should be mentioned, SparkSW method uses Spark version 1.0; however, the newest version used in our cluster is 2.0, which performs better in theory. Additionally, we deploy the newest Hadoop framework on our cluster for running HAlign.

Experiment (a). Based on MUSCLE [33], MAFFT [34], HAlign, and HAlign-II tools, we implement ultra-large multiple similar genome sequence alignments with $\Phi_{DNA}(1\times)$, $\Phi_{DNA}(100\times)$, and $\Phi_{DNA}(1000\times)$ datasets.

Experiment (b). Based on MUSCLE, MAFFT, HAlign, and HAlign-II tools, we implement ultra-large multiple dissimilarity RNA sequence alignments with $\Phi_{RNA}$(small) and $\Phi_{RNA}$(large) datasets.

Experiment (c). Based on MUSCLE, MAFFT, SparkSW, and HAlign-II tools, we implement ultra-large multiple dissimilarity protein sequence alignments with $\Phi_{Protein}$(1×), $\Phi_{Protein}$(100×), and $\Phi_{Protein}$(1000 ×) datasets.

Experiment (d). Based on IQ-TREE [35], HPTree, and HAlign-II tools, we construct ultra-large phylogenetic trees with $\Phi_{DNA}$(1×), $\Phi_{DNA}$(100×), $\Phi_{DNA}$(1000×), $\Phi_{RNA}$(small), $\Phi_{RNA}$(large), $\Phi_{Protein}$(1×), $\Phi_{Protein}$(100×), and $\Phi_{Protein}$(1000 ×) datasets. For our HAlign-II method, we initially align multiple sequences and then build phylogenetic trees.

(Table 2)

(Table 3)

(Table 4)

(Table 5)

Tables 2, 3, and 4 respectively show all experiment results with genome MSA, RNA MSA, and protein MSA. Surprisingly, MUSCEL exhibits extreme time consumption. Based on our experiments, MUSCEL performs best with small datasets, but it cannot properly allocate memory resource, resulting in high memory occupancy rate. Hence, MUSCEL eventually reports an out-of-memory message with ultra-large

datasets. Similarly, MAFFT proves to be incapable under such occasion. Based on Hadoop framework, HAlign and HPTree perform better, but many key-value pair conversion operators also result in high memory occupancy rate. Considering the problems leading to degraded performance, HAlign-II utilizes memory operation on hard disks, cutting down space complexity and memory occupancy rate. These improvements facilitate running of sequence analysis on clusters comprising cheap large-scale and low-end machines. However, HAlign-II features an average SP score that is inferior to those of other methods. Our method ignores high precision for changing large-scale computing power, which is necessary for several decision research.

Table 5 presents running times of several outstanding tools on phylogenetic trees construction. IQ-TREE with multiple threads consumes more time than HPTree and HAlign-II, as distributed computing on a single node utilizes multiple threads and features time-efficient data construction. Phylogenetic tree performance is evaluated by maximum likelihood value under log functions. HPTree point reaches -21954385, which is similar to that of NJ model in MEGA [36], implying close performance of results of both methods. Similarly, out-of-memory error occurs when running the HPTree method. Currently, no outstanding method exists for constructing large-scale evolutionary trees, even on workstation clusters. Constructing phylogenetic trees based on MSA results can speed up construction speed.

**Memory efficiency and scalability**

Currently, most time-efficient methods, such as MUSCEL with small datasets and HAlign for large-scale datasets, present extremely large space complexities, resulting in impossibility to actually address ultra-large datasets. Based on $\Phi_{DNA}(100\times)$, $\Phi_{DNA}(1000\times)$, $\Phi_{Protein}(100\times)$, and $\Phi_{Protein}(1000\times)$ datasets, HAlign for genome MSA and SparkSW for protein MSA are compared to describe memory efficiency of HAlign-II. We also design another comparison experiment to demonstrate scalability of HAlign-II.

(Figure 5)

(Figure 6)

Figure 5 shows average maximum memory usage of each machine on the cluster containing 12 machines. To conclude, Spark framework exhibits more efficient memory than Hadoop framework, as shown by inferiority of HAlign compared with other methods. Whether for nucleotide sequences or protein sequences, HAlign-II presents the lowest average maximum memory usage, thereby facilitating ultra-large MSA and phylogenetic tree construction on cheaper clusters. Additionally, Figure 6 shows that with increase in worker nodes, running time and memory efficiency becomes significantly low, indicating linear growth of capacity and computing power with increase of such nodes.

## Discussion

Multiple biological sequence alignment and phylogenetic tree construction present complicated inter-relationships, and both are necessary for sequence analysis. In the last several decades, many state-of-the-art methods and algorithms were created for more time- and space-efficient MSA and phylogenetic trees construction issues. With increasing next-generation sequence database, addressing ultra-large datasets became an unprecedented challenge. Other outstanding methods were developed to improve time efficiency even with precision loss; such methods include ClustalW-MPI, Hadoop-BAM, HAlign, and HPTree. Thus, with the urgent need for additional time-efficient and computing power for ultra-large datasets, we conduct a series of experiments to assess the performance of our HAlign-II method.

Based on Spark distributed and parallel computing model, Smith–Waterman algorithm, trie trees, and NJ methods are employed to completely utilize hardware resources and computing power. For ultra-large genome and RNA MSA experiments, MUSCEL and MAFFT achieve high accuracies. However, both traditional tools show complete incompatibility with large datasets. Methods based on distributed computing model present remarkable advantages, especially HAlign-II, which presents the highest memory efficiency. SparkSW and HAlign-II work well for ultra-large protein MSA experiments. However, the former still needs to further cut down memory occupation. Difficulty also arises from insufficient phylogenetic tree construction for ultra-large protein sequences. For ultra-large phylogenetic tree construction based on MSA results, most tools run out of memory, and even nearly 400 GB memory cannot address the requirement of 10 GB size datasets. All experimental results indicate that with regard to ultra-large nucleotide MSA or protein MSA and phylogenetic tree construction, HAlign-II performs best with regard to time efficiency, memory efficiency, and scalability.

## Conclusion

This paper implements a distributed and parallel computing tool named HAlign-II to address ultra-large multiple biological sequence alignment and phylogenetic tree construction. After comparing this tool with a series of state-of-the-art methods with ultra-large data, we conclude that HAlign-II features three advantages: 1) extremely high memory efficiency and good scaling with increases in computing resource; 2) efficient construction of phylogenetic trees with ultra-large biological sequences; 3) provision of user-friendly web server based on high performance and distributed computing infrastructure; the server is established at http://lab.malab.cn/soft/halign. These improvements will be significant in coping with extreme increases in next-generation sequencing.

## Ethics approval and consent to participate

None.

## Consent for publication

Agree.

## Availability of data and material

Available.

## Competing interests

None.

## Funding

The work was supported by the Natural Science Foundation of China (No. 61370010).

## Authors' contributions

Mr. Shixiang Wan and Prof. Quan Zou conceived and designed the study. Mr. Shixiang Wan performed the experiments and wrote the paper. Prof. reviewed and edited the manuscript. All authors read and approved the manuscript.

## Acknowledgments

The work was supported by the Natural Science Foundation of China (No. 61370010).

# Figures

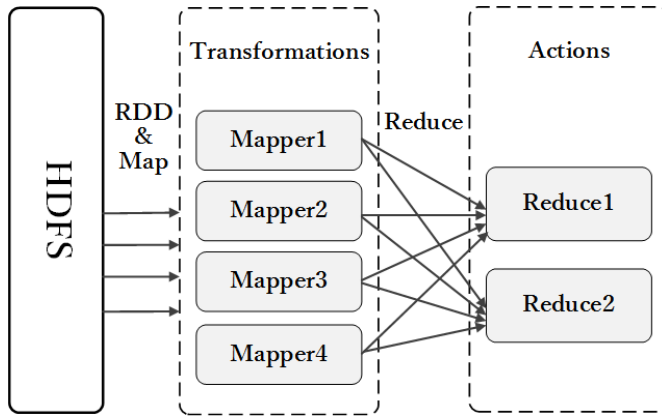

Figure 1. A simple Spark workflow.

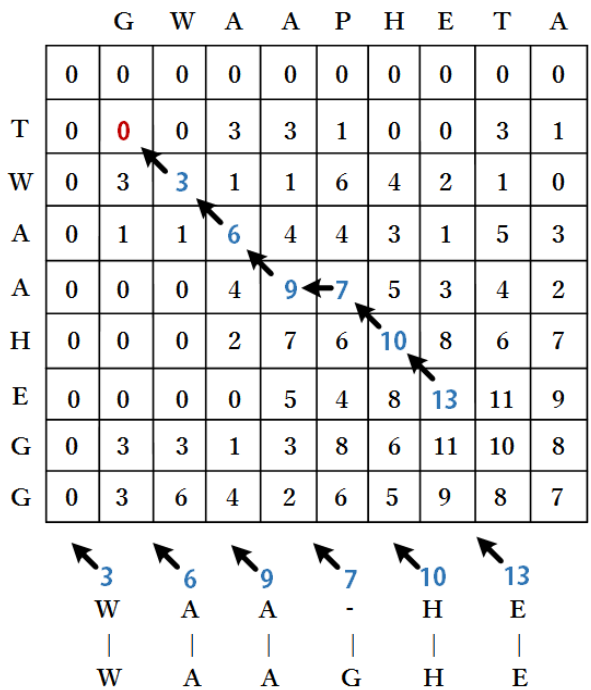

Figure 2. Traceback procedure and pairwise alignment results of Smith-Waterman algorithm.

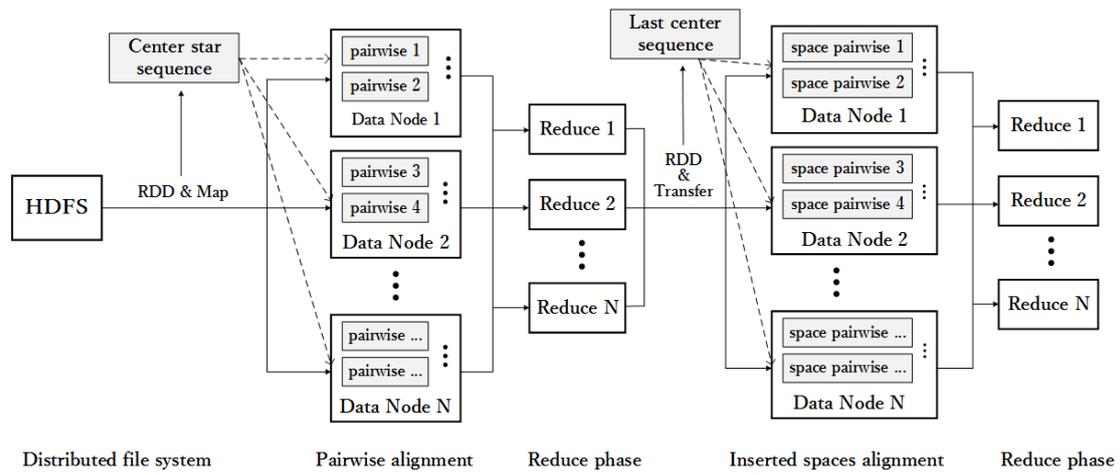

**Figure 3. MSA procedures based on Spark distributed framework.**

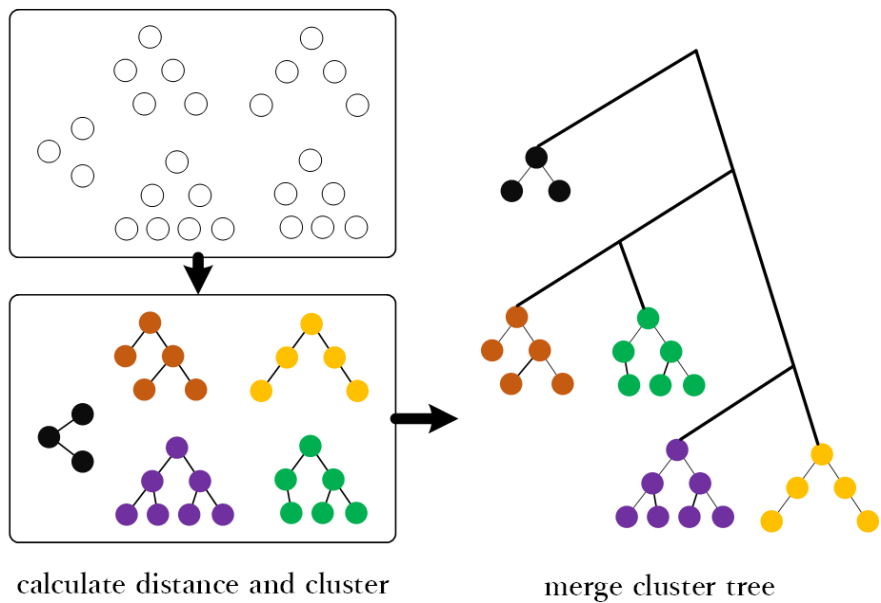

**Figure 4. Constructing phylogenetic trees based on distance measure.**

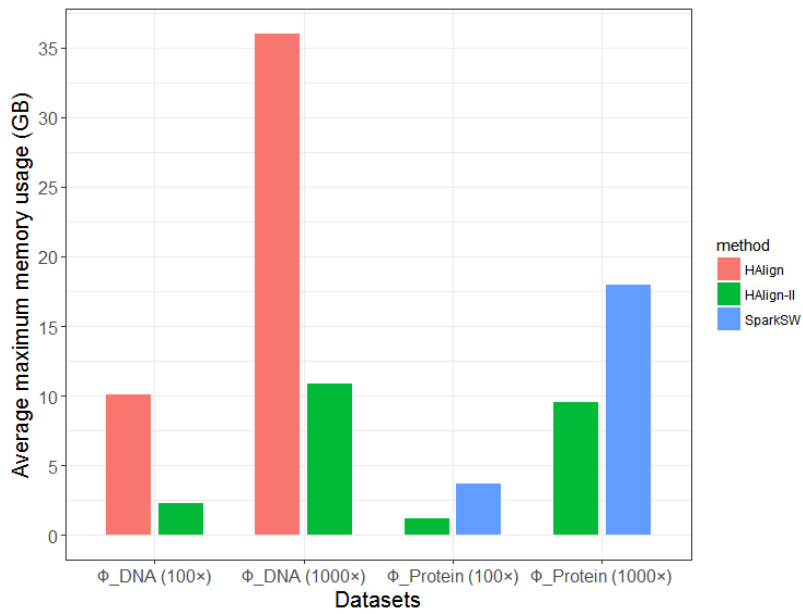

**Figure 5. Average maximum memory usage of protein MSA on clusters.**

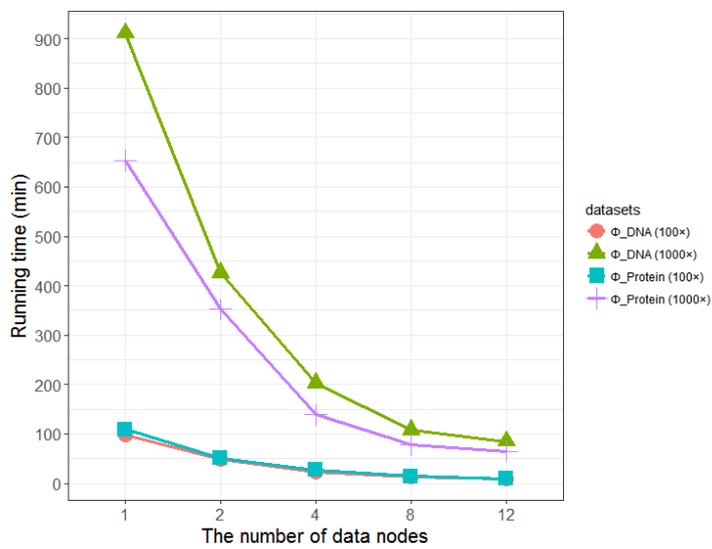

**Figure 6. Running time with increasing worker nodes.**

# Tables

**Table 1. Original dataset and datasets after threshold removal.**

| Dataset | Number | Minimum length | Maximum length | Average length | File size |
|---|---|---|---|---|---|
| $\Phi_{DNA}(1\times)$ | 672 | 16556 | 16579 | 16569.7 | 10 MB |
| $\Phi_{DNA}(100\times)$ | 67200 | as above | as above | as above | 1.1 GB |
| $\Phi_{DNA}(1000\times)$ | 672000 | as above | as above | as above | 11 GB |
| $\Phi_{RNA}(small)$ | 108453 | 807 | 1599 | 1442.8 | 156 MB |
| $\Phi_{RNA}(large)$ | 1011621 | 807 | 1629 | 1388.5 | 1.4 GB |
| $\Phi_{Protein}(1\times)$ | 17892 | 19 | 4895 | 459.0 | 15 MB |
| $\Phi_{Protein}(100\times)$ | 1789200 | as above | as above | as above | 1.5 GB |
| $\Phi_{Protein}(1000\times)$ | 17892000 | as above | as above | as above | 15 GB |

**Table 2. Running time and average SP values with genome MSA.**

|  | $\Phi_{DNA}(1\times)$ | | $\Phi_{DNA}(100\times)$ | | $\Phi_{DNA}(1000\times)$ | |
|---|---|---|---|---|---|---|
|  | time | avg SP | time | avg SP | time | avg SP |
| MUSCLE | 6 h 15 m | 81 | - | - | - | - |
| MAFFT | 1 m 20 s | 152 | - | - | - | - |
| HAlign | 2 m 12 s | 191 | 26 m 35 s | 191 | 5 h 28 m | 191 |
| HAlign-II | 14 s | 195 | 10 m 24 s | 195 | 1 h 25 m | 195 |

**Table 3. Running time and average SP values with RNA MSA.**

|  | $\Phi_{RNA}(small)$ | | $\Phi_{RNA}(large)$ | |
|---|---|---|---|---|
|  | Time | avg SP | Time | avg SP |
| MUSCLE | - | - | - | - |
| MAFFT | > 24 h | 26743 | - | - |
| HAlign | 1 h 32 s | 15660 | 3 h 15 m | 32079 |
| HAlign-II | 23 m 34 s | 16620 | 59 m 42 s | 35956 |

**Table 4. Running time and average SP values with protein MSA.**

|  | $\Phi_{Protein}(1\times)$ | | $\Phi_{Protein}(100\times)$ | | $\Phi_{Protein}(1000\times)$ | |
|---|---|---|---|---|---|---|
|  | time | avg SP | time | avg SP | time | avg SP |
| MUSCLE | - | - | - | - | - | - |
| MAFFT | 5 m 34 s | 925 | - | - | - | - |
| SparkSW | 1 m 56 s | 1009 | 50 m 51 s | 1009 | 4 h 34 m | 1009 |
| HAlign-II | 30 s | 1131 | 10 m 12 s | 1131 | 1 h 5 m | 1131 |

**Table 5. Running time during phylogenetic trees construction.**

|  | IQ-TREE | HPTree | HAlign-II |
|---|---|---|---|
| $\Phi_{DNA}(1\times)$ | 9 m 52 s | 1 m 25 s | 27 s |
| $\Phi_{DNA}(100\times)$ | 1 h 2 m | 45 m 32 s | 17 m 45 s |
| $\Phi_{DNA}(1000 \times)$ | - | - | 1 h 45 m |
| $\Phi_{RNA}(small)$ | - | 6 h 23 m | 52 m 39 s |
| $\Phi_{RNA}(large)$ | - | > 24 h | 8 h 20 m |
| $\Phi_{Protein}(1\times)$ | 13 m 26 s | not supported | 35 s |
| $\Phi_{Protein}(100\times)$ | 1 h 47 m | not supported | 15 m 23 s |
| $\Phi_{Protein}(1000 \times)$ | - | not supported | 1 h 27 m |